\documentclass[11pt,a4paper]{article}
 \pdfoutput=1
\usepackage{jheppub_kim}
\usepackage{rotating}
\usepackage{graphicx,epsfig}
\usepackage{amsmath}
\usepackage {amssymb}
\usepackage{subfigure}

 \usepackage[utf8x]{inputenc}

\usepackage{relsize}

\usepackage{graphicx}
\usepackage{epsfig}
\usepackage{bm}
\usepackage{amsfonts}
\usepackage{amssymb}
\usepackage{color}
\usepackage{float}
\usepackage{amsmath}
\usepackage{dcolumn}
\usepackage{hyperref}
\usepackage{array}
\usepackage{lscape}

\usepackage{array,multirow}
\usepackage{soul}
\usepackage{subfigure}
\usepackage{dsfont}
\usepackage{hyperref}
\usepackage{txfonts}
\usepackage{newlfont}
\usepackage{times}

\providecommand{\U}[1]{\protect\rule{.1in}{.1in}}

\def\be{\begin{equation}}
\def\ee{\end{equation}}
\def\bea{\begin{eqnarray}}
\def\eea{\end{eqnarray}}
\def\eqi{\begin{equation}}
\def\eqf{\end{equation}}
\def\eqia{\begin{eqnarray}}
\def\eqfa{\end{eqnarray}}

\title{Observational constraints on Myrzakulov gravity}

 \author[a]{Fotios K. Anagnostopoulos}

\author[b,c]{Spyros Basilakos}

\author[c,d,e]{Emmanuel N. Saridakis}

\affiliation[a]{Department of Physics, National \& Kapodistrian University of 
Athens, 
Zografou Campus GR 157 73, Athens, Greece}

 \affiliation[b]{Academy of Athens, Research Center for Astronomy and
Applied Mathematics, Soranou Efesiou 4, 11527, Athens, Greece}

\affiliation[c]{National Observatory of Athens, Lofos Nymfon, 11852 Athens, 
Greece}

  \affiliation[d]{CAS Key Laboratory for Researches in Galaxies and Cosmology, 
Department of Astronomy, University of Science and Technology of China, Hefei, 
Anhui 230026, P.R. China}
 \affiliation[e]{School of Astronomy, School of Physical Sciences, 
University of Science and Technology of China, Hefei 230026, P.R. China}

\emailAdd{fotis-anagnostopoulos@hotmail.com}
\emailAdd{svasil@academyofathens.gr} 
\emailAdd{msaridak@noa.gr}

\abstract{  
 We use data from Supernovae (SNIa) Pantheon sample, from Baryonic Acoustic 
Oscillations (BAO), and from cosmic chronometers measurements of the Hubble 
parameter (CC), alongside arguments from Big Bang Nucleosynthesis (BBN), in 
order to extract constraints on  Myrzakulov $F(R,T)$ gravity. This is a 
connection-based theory belonging to the Riemann-Cartan subclass, 
that uses a specific but non-special connection, which then leads to  extra 
degrees of freedom. Our analysis shows that both considered models lead to 
$\sim 1 \sigma$ compatibility in all cases. For the involved dimensionless 
parameter we find that it is constrained to an interval around 
zero, however the corresponding contours are slightly shifted towards positive 
values.
 Furthermore, we use the obtained 
parameter chains  so to reconstruct the corresponding Hubble function, as well 
as the
dark-energy 
equation-of-state parameter, as a function of  redshift. As we show,  
Model 1 is very close to $\Lambda$CDM scenario, while    Model 2 
  resembles  it at low redshifts, however at earlier times
 deviations  are  allowed.
  Finally, applying the AIC, BIC and the combined DIC criteria, we 
deduce that both models  present a very efficient fitting behavior, and 
are statistically equivalent with $\Lambda$CDM cosmology, despite the fact that 
Model 2 does not contain the latter as a limit. 
  
}
\keywords{Modified gravity, Observational Constraints, Torsional Gravity}

\begin{document}
\maketitle

\section{Introduction}

  According to the concordance cosmological model    the universe 
experienced 
two epochs of accelerated expansion, one at early and one at late times. 
Although the latter can be explained by the presence of a cosmological constant,
the related theoretical problem, the possibility of a dynamical behavior, and 
especially the inability of the cosmological constant to describe the early 
accelerated phase, led to the incorporation of some form of modification. 
As a first possibility one can maintain general relativity as the underlying 
theory and modify the matter content of the universe by introducing  extra  
fields, 
such as the 
inflaton  at early times  
\cite{Olive:1989nu,Bartolo:2004if}  and/or 
the dark energy sector at late times  
\cite{Copeland:2006wr,Cai:2009zp}.
As a second possibility one modifies the gravitational sector itself, 
constructing a theory that  possesses general 
relativity 
as a particular limit but which in general  exhibits
extra degrees of freedom  \cite{Capozziello:2011et,Cai:2015emx}.  

There are many ways to construct gravitational modifications, each one 
modifying a particular feature of general relativity. Modifying the 
dimensionality  gives rise to the braneworld theories \cite{Brax:2004xh}, 
modifying the 
  Einstein-Hilbert Lagrangian, gives rise to 
  $F(R)$ gravity 
\cite{DeFelice:2010aj,Nojiri:2010wj},  $F(G)$ gravity   
\cite{Nojiri:2005jg,DeFelice:2008wz},   Lovelock theories
\cite{Lovelock:1971yv,Deruelle:1989fj}, etc, while adding a scalar field 
coupled with curvature in various ways gives rise to Horndeski/Galileon 
theories \cite{Horndeski:1974wa,Nicolis:2008in,Deffayet:2009wt}. Additionally, 
starting from the equivalent, teleparallel, formulation of gravity
\cite{Pereira,Maluf:2013gaa} one can construct modifications using torsional 
invariants, such as in  $F(T)$ gravity 
\cite{Ferraro:2006jd,Linder:2010py}, in $F(T,T_G)$ 
gravity \cite{Kofinas:2014owa}, or in scalar-torsion 
theories \cite{Geng:2011aj,Hohmann:2018rwf}.
Moreover, one can construct the general class of metric-affine 
theories \cite{Hehl:1994ue,BeltranJimenez:2012sz,Tamanini:2012mi}, 
  which incorporates a general  linear  connection structure, or proceed to the 
introduction of  non-linear connections such as in  
in Finsler and Finsler-like theories  
\cite{Bogoslovsky:1999pp,Mavromatos:2010jt,Kouretsis:2012ys,Basilakos:2013hua,
Triantafyllopoulos:2018bli,Ikeda:2019ckp}.

Inspired by these, one could start from such affinely connected metric 
theories, 
and in particular from their  Riemann-Cartan subclass \cite{Conroy:2017yln}, 
and construct a theory using a specific but 
non-special connection, which would lead to  non-zero torsion and non-zero 
curvature at the same time, and thus offering the extra degrees of freedom 
typically needed in any gravitational modification \cite{Myrzakulov:2012qp}.
Myrzakulov   gravity can thus lead to a good  phenomenology, being able to 
describe the universe evolution at early and late times 
\cite{Saridakis:2019qwt,Sharif:2012gz,Momeni:2011am,
Capozziello:2014bna,
Feola:2019zqg}. 

One basic question in    modified gravities is the determination of 
the arbitrary 
function that enters in the theory. Although some general features can be 
deduced 
through theoretical considerations, such as the absence of ghosts and
 instabilities, or the existence of Noether symmetries,  the most powerful tool 
is 
the use of   observational data  
\cite{Feng:2004ad,Olivares:2005tb,Capozziello:2005pa,Maartens:2006yt,
Lazkoz:2006gp,WoodVasey:2007jb,Zhang:2007sh,Tsujikawa:2007xu,
Basilakos:2009wi,Dutta:2010jh,Nesseris:2010pc,Geng:2011ka,Basilakos:2013nfa,
Basilakos:2015vra,Yang:2018qmz,Anagnostopoulos:2019miu,Pan:2019gop,
Anagnostopoulos:2020ctz,Alfaro:2020dmb,Felegary:2020zuc,
DiValentino:2020vnx}. Hence, in this work we are 
interested in using 
 expansion data  such as  Supernovae type Ia
data  (SNIa),    
Baryonic Acoustic Oscillations (BAO), and Hubble Cosmic Chronometers (CC) 
observations, in order to impose constraints on Myrzakulov   gravity.
The plan of the work is the following. In Section \ref{section2} we present  
Myrzakulov gravity and its cosmological 
applications. In Section 
\ref{DataMethodology} we describe the various datasets and  the involved 
statistical 
methods. Then in Section 
\ref{results} we preform our analysis and we present the results, namely the 
constraints on the various parameters.
Finally, in  Section 
 \ref{Conclusions} we summarize and conclude.

\section{Myrzakulov Gravity and Cosmology}
\label{section2}

In this section we present a brief review of Myrzakulov gravity, or $F(R,T)$ 
gravity
\cite{Myrzakulov:2012qp,Saridakis:2019qwt}, extracting additionally the 
relevant cosmological equations.

\subsection{Myrzakulov Gravity}

The central idea of this modified gravity is 
the modification of the underlying connection. In particular, it is known that 
imposing a general connection $\omega^{a}_{\,\,\,bc}$ one defines the 
 curvature and the torsion tensor respectively 
 as \cite{Kofinas:2014owa}
\begin{eqnarray}
&&\!\!\!\!\!\!\!\!\! R^{a}_{\,\,\, b\mu\nu}\!=\omega^{a}_{\,\,\,b\nu,\mu}-
\omega^{a}_{\,\,\,b\mu,\nu}
+\omega^{a}_{\,\,\,c\mu}\omega^{c}_{\,\,\,b\nu}-\omega^{a}_{\,\,\,c\nu}
\omega^{c}_{\,\,\,b\mu}\,,
\label{curvaturebastard}
\end{eqnarray}
\begin{equation}
T^{a}_{\,\,\,\mu\nu}=
e^{a}_{\,\,\,\nu,\mu}-e^{a}_{\,\,\,\mu,\nu}+\omega^{a}_{\,\,\,b\mu}e^{b}_{\,\,\,
\nu}
-\omega^{a}_{\,\,\,b\nu}e^{b}_{\,\,\,\mu}\,,
\label{torsionbastard}
\end{equation}
where $e^{\,\,\, \mu}_a\partial_\mu$ is the tetrad field related to the metric 
through  $
g_{\mu\nu} =\eta_{ab}\, e^a_{\,\,\,\mu}  \, e^b_{\,\,\,\nu},
$ where $\ \eta_{ab}=\text{diag}(1,-1,-1,-1)$, with Greek and Latin indices 
denoting coordinate and tangent space respectively, and where 
comma denotes differentiation.

There are infinite connection choices. The    Levi-Civita     
$\Gamma_{abc}$ is the only connection that gives 
 vanishing torsion, and from now on we use the label ``LC'' to denote the 
 curvature (Riemann) tensor corresponding to $\Gamma_{abc}$, namely
 $R^{(LC)a}_{\,\,\,\ \ \ \ \ \, 
b\mu\nu}=\Gamma^{a}_{\,\,\,b\nu,\mu}-
\Gamma^{a}_{\,\,\,b\mu,\nu}
+\Gamma^{a}_{\,\,\,c\mu}\Gamma^{c}_{\,\,\,b\nu}-\Gamma^{a}_{\,\,\,c\nu}
\Gamma^{c}_{\,\,\,b\mu}$.
On the other hand one can use the Weitzenb{\"{o}}ck connection  
$W_{\,\,\,\mu\nu}^{\lambda}=e_{a}^{\,\,\,\lambda}e^{a}_{\,\,\,\mu
, \nu } $ which is curvatureless, leading only to torsion as
 $T^{(W)\lambda}_{\,\,\,\ \ \ \ \ \mu\nu}=W^{\lambda}_{\,\,\,\nu\mu}-
W^{\lambda}_{\,\,\,\mu\nu}$
(we use the label ``W'' to denote quantities corresponding to 
$W_{\,\,\,\mu\nu}^{\lambda}$.
 From the above it is implied that the Ricci scalar corresponding to the 
Levi-Civita connection is 
  \begin{eqnarray}
 &&
 \!\!\!\!\!\!\!\!\!\!\!\!\!\!\!\!\!\!\!\!\!\!\!\!\!\!\!\!\!\!
 R^{(LC)}=\eta^{ab} e^{\,\,\, \mu}_a e^{\,\,\, \nu}_b  \left[
 \Gamma^{\lambda}_{\,\,\,\mu\nu,\lambda}
-
 \Gamma^{\lambda}_{\,\,\,\mu\lambda,\nu} 
 + \Gamma^{\rho}_{\,\,\,\mu\nu}\Gamma^{\lambda}_{\,\,\,\lambda\rho}
-\Gamma^{\rho}_{\,\,\,\mu\lambda}\Gamma^{\lambda}_{\,\,\,\nu\rho}
  \right],
\end{eqnarray}
while the torsion scalar corresponding to the Weitzenb{\"{o}}ck connection 
is
\begin{eqnarray} 
&&
\!\!\!\!\!\!\!\!\!\!
T^{(W)}=\frac{1}{4}
\left(W^{\mu\lambda\nu}-
W^{\mu\nu\lambda} \right)
\left(W_{\mu\lambda\nu}  -W_{\mu\nu\lambda}\right) 
+\frac{1}{2} \left(W^{
\mu\lambda\nu }
-W^{
\mu\nu\lambda } \right)
\left(W_{\lambda\mu\nu}
-W_{\lambda\nu\mu}\right)
\nonumber\\
&&\ \ \ \ \ 
- \left(  
W_{\nu}^{\,\,\,\mu\nu}
-W_{\nu}^{\,\,\,\nu\mu}\right)  
\left( W^{\lambda}_{\,\,\,\mu\lambda}-W^{\lambda}_{\,\,\,\lambda\mu}\right).
\label{TdefW}
 \end{eqnarray}
 As it is known, the former is used in the Lagrangian of General Relativity and 
in all curvature-based modified gravities, e.g. in  $F(R)$ gravity 
\cite{DeFelice:2010aj}, while the latter is used in the Lagrangian 
of teleparallel equivalent of general relativity and in all torsion-modified 
gravities, e.g. in $F(T)$ gravity  \cite{Cai:2015emx}.

In  Myrzakulov   gravity one uses a non-special  connection which has  non-zero 
curvature and torsion simultaneously \cite{Saridakis:2019qwt}. Hence,   the 
resulting theory will in general 
possess extra degrees of freedom, even if the imposed Lagrangian is simple, 
which is not the case of general relativity or of   teleparallel equivalent 
of general relativity that both have two degrees of freedom corresponding to 
the 
massless graviton. The
  action of the theory is  
\begin{equation}
S = \int d^{4}x e \left[ \frac{F(R,T)}{2\kappa^{2}}   +L_m \right],
\label{action1}
\end{equation}
where $e = \text{det}(e_{\mu}^a) = \sqrt{-g}$,  $\kappa^2=8\pi G$ is  the 
gravitational constant, and where  we have   introduced 
the matter Lagrangian $L_m$ too for completeness. Note that   in 
the   arbitrary function  $F(R,T)$  the $R$ and $T$ are the curvature and 
torsion scalars corresponding 
to the   non-special connection used, which read as   
\cite{Kofinas:2014owa}
\begin{eqnarray}
&&
 T=\frac{1}{4}T^{\mu\nu\lambda}T_{\mu\nu\lambda}+\frac{1}{2}T^{\mu\nu\lambda}
T_{\lambda\nu\mu}-T_{\nu}^{\,\,\,\nu\mu}T^{\lambda}_{\,\,\,\lambda\mu},
\label{Tdef2}
\\
&&
 R=R^{(LC)}+T-2T_{\nu\,\,\,\,\,\,\,\,;\mu}^{\,\,\,\nu\mu}\,,
 \label{Radef222}
 \end{eqnarray}
where $;$ marks the covariant differentiation with respect to the 
Levi-Civita connection. Therefore,  $T$ depends on the tetrad field, its first 
derivative and the connection, while $R$ depends on the tetrad and its first 
derivative, and on the connection and its first derivative, with an additional 
dependence on the second tetrad derivative due to the last term 
of (\ref{Radef222}).  
Thus, using (\ref{TdefW}),(\ref{Tdef2}),(\ref{Radef222}) 
we can finally write  
\begin{eqnarray}
&&T=T^{(W)}+v, 
\label{T1}
\\
&&
R=R^{(LC)} + u, 
\label{R1}
\end{eqnarray}
where $v$ is a scalar depending   on the tetrad, its first 
derivative and the connection, while $u$ is a scalar   depending on 
the   tetrad, its first and second derivatives, and the connection 
and its first derivative.

The quantities  $u$ and $v$  quantify the effect of the 
specific but non-special imposed connection. 
 In the case where this connection becomes the Levi-Civita 
one, then $u=0$ and $v=-T^{(W)}$, and the above theory becomes   the 
usual $F(R)$ gravity, which  in turn coincides with general relativity under 
$F(R)=R$. 
On the other hand, in the case where the   connection is the 
Weitzenb{\"{o}}ck 
one, then we have that $v=0$ and $u=-R^{(LC)}$ and hence the theory coincides 
with $F(T)$ 
gravity,  which in turn becomes the teleparallel equivalent of general 
relativity for $F(T)=T$.

 \subsection{Cosmology}
  
Let us now apply the above into a cosmological framework and extract the 
corresponding equations that determine the universe evolution. As it was shown 
in \cite{Saridakis:2019qwt}, in order to avoid complications related to the 
additional variation in terms of the 
connection, it proves convenient to  apply a mini-super-space procedure.
Hence, we apply  the homogeneous and isotropic  flat 
Friedmann-Robertson-Walker (FRW) geometry 
\begin{eqnarray}
 ds^2= dt^2-a^2(t)\,  \delta_{ij} dx^i dx^j,
 \end{eqnarray}
which corresponds to the tetrad   
$e^a_{\,\,\,\mu}={\text
diag}[1,a(t),a(t),a(t)]$,
where $a(t)$ is the scale factor.
In this case one can easily find  that
$R^{(LC)}=6   \left( \frac{\ddot{a}}{a}+  \frac{\dot{a}^{2}}{a^2}\right)$ and 
$T^{(W)}=-6 \left(  \frac{\dot{a}^{2}}{a^2} \right)$. Furthermore, we use the 
standard replacement $L_m=-\rho_m(a)$ 
\cite{Paliathanasis:2014iva,Paliathanasis:2015aos,Dimakis:2016mip}.
Lastly, following the discussion on the dependence of $u$ and $v$ above, we 
consistently impose that  $u=u(a,\dot{a},\ddot{a})$ and $v=v(a,\dot{a})$.

 In this work we are interested in exploring 
 the cosmological behavior that arise purely from the non-special connection of 
Myrzakulov gravity. Hence, we focus on the  
 simplest case where the involved arbitrary function is trivial, namely  
 $F(R,T)=R+\lambda T$ with $\lambda$ a dimensionless parameter (we omit the 
coupling coefficient of $R$ since it
can be absorbed into $\kappa^2$). Note that we do not consider an explicit 
cosmological constant term in the Lagrangian. Inserting the above 
mini-super-space expressions  
into the action (\ref{action1}), for this Lagrangian choice  we acquire $S=\int 
Ldt$, where 
\begin{eqnarray}
&&
\!\!\!\!\!\!\!\!\!\!\!
L= 
\frac{3}{\kappa^2}\left[\lambda+1\right]a\dot{a}^{2}-
\frac{ a^{3}}{2\kappa^2}\left[ u(a,\dot{a},\ddot{a})+\lambda
v(a,\dot{a}) \right] 
+ a^3  \rho_m(a)
.\label{4.2}
\end{eqnarray}
Extracting the equations of motion for $a$, alongside   
the Hamiltonian constraint  $
{\cal H}=\dot{a}\left[\frac{\partial L}{\partial 
\dot{a}}- \frac{\partial}{\partial t}\frac{\partial L}{\partial 
\ddot{a}}\right]+\ddot{a}\left(\frac{\partial L}{\partial 
\ddot{a}}\right)-L=0 $, we finally acquire the Friedmann 
equations 
\begin{eqnarray}
3H^{2}&=& \kappa^2\left( 
\rho_m+\rho_{de} \right)
\label{FR1a}
\\
2\dot{H}+3H^2
&=&  -\kappa^2 \left(p_m+ p_{de}\right),
\label{FR2a}
\end{eqnarray}
where 
\begin{eqnarray}
&&
\!\!\!\!\!\!\!\!\!\!\!\!\!\!\!\!
\rho_{de}=\frac{1}{\kappa^2} \left[
\frac{Ha}{2} \left(u_{\dot{a}}+v_{\dot{a}} \lambda\right) -\frac{1}{2} 
(u+\lambda v) 
+
\frac{a u_{\ddot{a}}}{2}  \left(\dot{H}-2 H^2\right)    
-3\lambda H^2\right]
\label{rhoDEa1}\\
&&
\!\!\!\!\!\!\!\!\!\!\!\!\!\!\!\!
p_{de}=
-\frac{1}{\kappa^2}
\left[\frac{Ha}{2} \left(u_{\dot{a}}+v_{\dot{a}} 
\lambda\right)
 -\frac{1}{2} (u+ \lambda v)
 -\frac{a}{6} 
\left(u_a+\lambda v_a-{\dot{u}_{\dot{a}}}-\lambda 
{\dot{v}_{\dot{a}}}\right)\right.\nonumber\\
&& \left. \ \  \ \  \ \  \ \,  \ 
-\frac{a}{2}\left(\dot{H}+3H^2\right)u_{\ddot{a}}-H a 
\dot{u}_{\ddot{a}}
-\frac{a}{6} 
\ddot{u}_{\ddot{a}}
-\lambda(2\dot{H}+3H^2)\right],
\label{pDEa1}
\end{eqnarray}
with $H=\frac{\dot{a}}{a}$   the Hubble parameter,  $p_m$   the matter 
pressure, and with the subscripts 
$a,\dot{a},\ddot{a}$ denoting partial derivatives with respect to this 
argument. 
Hence, in the theory at hand,
 we 
obtain an effective dark energy sector which  arises from the    non-special 
connection. Additionally,   given the matter conservation equation 
$ \dot{\rho}_m+3H(\rho_m+p_m)=0$
we find
\begin{eqnarray}
 \dot{\rho}_{de}+3H(\rho_{de}+p_{de})=0,
\end{eqnarray}
which implies that 
  the effective dark energy  sector is conserved.  

 The above    Friedmann equations can efficiently describe the 
  late-time acceleration. A first observation is that   in the case 
where $\lambda=0$, namely in the case where the Lagrangian is just  the 
curvature (nevertheless the non-special connection leads to non-zero torsion 
too), and for the choice   $u=c_1 \dot{a}-c_2$, with $c_1$,$c_2$ constants, 
then 
we have $
 \rho_{de}=-p_{de}=\frac{c_2}{2\kappa^2}\equiv\Lambda$. Hence, the scenario at 
hand includes $\Lambda$CDM cosmology as a sub-case, although we have not 
considered 
an explicit cosmological constant, since the cosmological constant arises 
effectively  from the connection structure of the   theory. Thus, we expect 
that a realistic model would be a deviation from the above  scenario. 

Finally, it proves convenient to introduce the  decceleration parameter as
\begin{equation}
    \label{deccel_func_def}
    q = -1 - \frac{\dot{H}}{H^2},
\end{equation}
which quantifies the cosmic acceleration. Defining additionally the density 
parameters $\Omega_m=\kappa^2\rho_m/(3H^2)$ and 
$\Omega_{de}=\kappa^2\rho_{de}/(3H^2)$, as well as the equation-of-state 
parameters $w_m\equiv p_m/\rho_m$ and $w_{de}\equiv p_{de}/\rho_{de}$, we can 
extract the useful expression
\begin{equation}
    \label{deccel_func_def}
    \frac{2q-1}{3} = \Omega_m w_m+\Omega_{de}w_{de}.
\end{equation}
Hence, in the standard case of dust matter, namely for $w_m\approx0$, we obtain 
\begin{equation}
    \label{wde11}
 w_{de}= \frac{2q-1}{3(1-\Omega_m)}.
\end{equation}
This expression allows one to find the evolution of the dark energy 
equation-of-state 
parameter, knowing the solution of the Friedmann equations, or knowing the 
observable values of $H(z)$ (where $z$ is the redshift defined through 
$1+z=a_0/a$ setting the current value of the scale factor to $a_0=1$).

In the following we will focus to two models which can satisfy these features.

\subsubsection{Model 1}

  Choosing $u=c_1 \dot{a}-c_2$ 
and $v=c_3\dot{a}-c_4$,  with $c_3$,$c_4$ constants, we obtain
\begin{eqnarray}
3H^{2}&=& \kappa^2\left( 
\rho_m+\rho_{de} \right)
\label{FR1a1}
\\
2\dot{H}+3H^2
&=&  -\kappa^2 \left(p_m+ p_{de}\right),
\label{FR2a1}
\end{eqnarray}
with
\begin{eqnarray}
&&\rho_{de}=\frac{1}{\kappa^2} \left[c-3\lambda H^2\right]
\label{rhoDEa}
\\
&&
p_{de}=
-\frac{1}{\kappa^2}
\left[c-\lambda(2\dot{H}+3H^2)\right],
\label{pDEa}
\end{eqnarray}
where $c\equiv c_2+c_4$. Hence, in this scenario the geometrical sector 
constitutes an effective dark energy sector with the above energy density and 
pressure, and an equation-of-state parameter of the form
\begin{eqnarray}
&&w_{de}=-1+\frac{2\lambda \dot{H}}{c-3\lambda H^2}.
\label{wDEa}
\end{eqnarray}
Interestingly enough, we can see that $w_{de}$ can be both larger or smaller 
than -1, and thus the effective dark energy can be quintessence or phantom 
like. 

This model has two parameters, namely $c,\lambda$, but one of them can be 
eliminated using the present value of the matter density parameter 
$\Omega_{m0}$ (from now on the subscript ``0'' denotes the current value of a 
quantity), since (\ref{FR1a1}) at present gives:
 \begin{eqnarray}
1= \Omega_{m0}+ \frac{c}{3  H_0^2}-\lambda.
\label{FR2a1b}
\end{eqnarray}
Additionally, the deceleration parameter (\ref{deccel_func_def}), using 
(\ref{rhoDEa}),(\ref{pDEa}), becomes
\begin{equation}
\label{qz_myrz1}
    q(z) = -1+ \frac{2\Omega_{m0}(1+z)^3}{\Omega_{m0}(1+z)^3 + 2(1 + \lambda -  
\Omega_{m0})},
\end{equation}
and thus its value at present is 
\begin{equation}
    \label{q_0_myrz1}
    q_0 = -1 + \frac{3  \Omega_{m0}}{2(1+\lambda)}.
\end{equation}
Comparing with the corresponding value of $\Lambda$CDM scenario, namely 
$q^{\Lambda}_0 = -1+3\Omega_{m0}/2$, we verify 
     that for the special case of $\lambda = 0$ the two scenarios coincide, as 
mentioned above.
Finally, note that from relation (\ref{qz_myrz1}) we can calculate the 
transition redshit, namely
the redshift in which $q$ transits from posotive to negative and  we have the 
onset of acceleration, finding
\begin{eqnarray}
z_{tr}=-1+2^{1/3} (1+\lambda-\Omega_{m0})^{1/3} \Omega_{m0}^{-1/3}. 
\label{ztrmod1}
 \end{eqnarray}

\subsubsection{Model 2}

As a second example   let us consider a more general model with  $u=c_1 
\frac{\dot{a}}{a}\ln\dot{a}$ 
and $v=s(a)\dot{a}$, with $s(a)$ an arbitrary function. In this case 
(\ref{rhoDEa1}),(\ref{pDEa1}) give
\begin{eqnarray}
3H^{2}&=& \kappa^2\left( 
\rho_m+\rho_{de} \right)
\label{FR1a2}
\\
2\dot{H}+3H^2
&=&  -\kappa^2 \left(p_m+ p_{de}\right),
\label{FR2a2}
\end{eqnarray}
with
\begin{eqnarray}
&&\rho_{de}=\frac{1}{\kappa^2} \left[\frac{c_1}{2}H-3\lambda H^2\right]
\label{rhoDEb}
\\
&&
p_{de}=
-\frac{1}{\kappa^2}
\left[\frac{c_1}{2}H+\frac{c_1}{6}\frac{\dot{H}}{H}-\lambda(2\dot{H}+3H^2)\right
] ,
\label{pDEb}
\end{eqnarray}
 while
\begin{eqnarray}
&&w_{de} =-1+\frac{2\lambda 
\dot{H}-\frac{c_1}{6}\frac{\dot{H}}{H}}{\frac{c_1}{2}H-3\lambda H^2}.
\label{wDEb}
\end{eqnarray}
Similarly to the previous example, for this case too  $w_{de}$ can be 
quintessence-like or 
phantom-like.

This model has two parameters, namely $c_1,\lambda$, but one of them can be 
eliminated using $\Omega_{m0}$, since (\ref{FR1a1}) at present time leads 
to:
 \begin{eqnarray}
1= \Omega_{m0}+ \frac{c_1}{6  H_0}-\lambda.
\label{FR2a1b2}
\end{eqnarray}
The deceleration parameter (\ref{deccel_func_def}) becomes
\begin{equation}
\label{qz_myrz2}
    q(z) = -1 + \frac{\frac{3}{2} \Omega_{m0}(1+z)^3}{3(1-\Omega_{m0} + 
\lambda) 
+(1+\lambda)^{-1}\left[(1-\Omega_{m0}+\lambda)^2 + 
(1+\lambda)(1+z)^3\Omega_{m0}\right]^{1/2}}
\end{equation}
 and its current value is
 \begin{equation}
     \label{q_0_myrz2}
     q_0 = -1 + \frac{\frac{3}{2} \Omega_{m0}}{3(1-\Omega_{m0} + \lambda) 
+(1+\lambda)^{-1}\left[(1-\Omega_{m0}+\lambda)^2 + 
(1+\lambda)\Omega_{m0}\right]^{1/2}}.
 \end{equation}
Finally, from relation (\ref{qz_myrz2}) we can calculate the transition redshit 
as
\begin{eqnarray}
  &&z_{tr}=-1+\frac{6^{1/3}(1+\lambda)^{-1/3}\Omega_{m0}^{-1/3}}{3}\nonumber\\
  &&
 \cdot\Big\{   10+9 
\lambda(2+\lambda-\Omega_{m0})-9\Omega_{m0}  
\nonumber\\
&&
-
\sqrt{28-36\Omega_{m0}+9[3\lambda (2+\lambda)-4\lambda 
\Omega_{m0}+\Omega_{m0}^2]
}
\Big\}^{1/3}.
\label{ztrmod2}
\end{eqnarray}

\section{Data and Methodology}
\label{DataMethodology}

In this section we describe the various datasets that are going to be used 
in our analysis, and  also the involved 
statistical 
methods. 
In particular, we will use   data from direct measurements of 
the
Hubble parameter,   from Supernovae Type Ia (SNIa), and from Baryonic Acoustic 
Oscillations. Finally, we present various information criteria that offer 
information  on the quality of the fit.

\subsection{Cosmological probes}

\subsubsection{Direct measurements of the Hubble expansion}
 	
          From the latest $H(z)$ data set compilation available in Ref. 
\cite{YuRatra2018} 
we use  only  data obtained from  cosmic chronometers (CC). By using 
the
differential age of passive evolving galaxies  one can measure the Hubble rate 
directly (see e.g. Ref. 
\cite{Moresco:2018xdr} and references therein). These galaxies are massive 
galaxies  that evolve ``slowly" at certain intervals of the cosmic time, i.e 
with small fraction of ``new" stars. A striking advantage of the 
differential 
age of passive evolving galaxies is that the resulting measurement of the 
Hubble 
rate 
comes without any assumptions for the underlying cosmology, with the exception 
of imposed spatial flatness. Our study 
incorporates $N=31$ 
measurements of the Hubble expansion in the redshift range $0.07 \lesssim z 
\lesssim 2.0 $.
 
Here, the corresponding $\chi^2_{H}$ function reads
          \begin{equation}
          \chi^{2}_{H}\left(\phi^{\nu}\right)= 
\sum_{i=1}^{N}\frac{\left[H^{obs}_{i}-H_{th}(z_{i};\phi^{\nu})\right]^2}{\sigma_
{i}^2}\,,
          \end{equation}
          where $H^{obs}_{i}$ is the observed Hubble
rate at redshift $z_{i}$ and $\sigma_{i}$ the corresponding uncertainty, while 
$\phi^{\nu}$  is the   statistical vector that
contains the free parameters of the  examined model.

\subsubsection{Supernovae Type Ia}

The most common class of cosmological probes is the so-called ``standard" 
candles. The latter are   luminous extra-galactic 
astrophysical objects with observable features that are independent of the 
cosmic 
time. 
The 
most known standard candles and probably the most thoroughly studied are 
Supernovae Type Ia (SNIa). In our 
analysis we use the most recent SNIa  dataset available, i.e the binned 
Pantheon sample of Scolnic et. al. \cite{Scolnic:2017caz}. The full 
dataset is   approximated very efficiently with the binned  $N = 40$ 
data points belonging to the redshift interval $0.01 \lesssim z \lesssim 1.6$.
          The corresponding $\chi^2$ is
          \begin{equation}
          \chi^{2}_{SN Ia}\left(\phi^{\nu+1}\right)={
          \bold 
\mu}_{\text{SNIa}}\,
          {\bold C}_{\text{SNIa},\text{cov}}^{-1}\,{\bold 
\mu}_{\text{SNIa}}^{T}\,,
          \end{equation}
          where
          ${\bold 
\mu}_{\text{\text{SNIa}}}=\{\mu_{1}-\mu_{\text{th}}(z_{1},\phi^{\nu})\,,\,...\,,
\,
  \mu_{N}-\mu_{\text{th}}(z_{N},\phi^{\nu})\}$. The distance modulus reads as
  $\mu_{i} = \mu_{B,i}-\mathcal{M}$, with $\mu_{B,i}$   the apparent maximum
magnitude      for           redshift $z_{i}$. Here,   $\mathcal{M}$ 
is a hyper-parameter   \cite{Scolnic:2017caz}   that 
quantifies uncertainties 
of 
various 
origins, such as astrophysical ones, data-reduction pipeline, etc, and it is 
employed instead of the usage of $\alpha, \ \beta$ free parameters,  in the 
context of ``BEAMS with Bias Corrections''
method  \cite{Kessler:2016uwi}. The observed distance modulus is  compared with 
the 
theoretical one, i.e
  \begin{equation}
  \mu_{\text{th}} = 5\log\left(\frac{d_{L}(z;\phi^{\nu})}{\text{Mpc}}\right) + 
25\,,
 \end{equation}
with
 \begin{equation}
d_L(z;\phi^{\nu}) = c(1+z)\int_{0}^{z}\frac{dx}{H(x,\phi^{\nu})}
\end{equation}
  the luminosity distance for   flat FRW geometry. It must be noted 
that  $\mathcal{M}$ and the normalized Hubble constant $h$ are  
degenerate in light of   Pantheon dataset in an intrinsic way, as it is 
usual in standard candles. Therefore, one should jointly employ  other 
data-sets in order to obtain meaningful information regarding  the present 
value  $H_{0}$.

\subsubsection{Baryonic Acoustic Oscillations}

Baryonic Acoustic Oscillations refer to the imprint left by relativistic sound 
waves in the early
universe, providing an observable to the late-time large scale structure. 
The main idea is to measure the aforementioned scale at different times (i.e 
redshifts), and thus obtain $D_A(z)$ and $H(z)$. The
acoustic length scale corresponds to the co-moving distance that the sound 
waves could travel until  the recombination $z_{*}$ 
\cite{Eisenstein:1997ik}, namely
\begin{equation}
    \label{r_s_theor}
    r_d = \int_{z_{*}}^{\infty}\frac{c_{s}(z)}{H(z)}.
\end{equation}
For the concordance model, the sound speed,  $c_{s}$,  is given from an 
analytical expression. However, for the models considered here  there is not 
such an expression, therefore the scale $r_d$ will be addressed as a free 
parameter. 
Furthermore, distances of different objects along the line of sight  correspond 
to different redshifts and thus   depend on the combination $H(z)r_d$, while 
distances 
transverse to the line of sight are related with the combination $D_A (z)/r_d$. 

Employing large samples of tracers, (i.e galaxies), one can detect by 
statistical means the BAO peak, (for details see  \cite{Alam:2020sor} and 
references therein). In order to achieve this  it is required to impose an 
underlying cosmology, and hence the method is not model independent. 
However, the 
differences that may infiltrate at the final data products  are     
  much less than the statistical errors, and in most cases the data points are 
calibrated with the quantity $r_{d,fid}/r_{d}$. In this work we employ the BAOs 
data-set used by   \cite{Ryan:2019uor}, that consists of $N = 11$ data 
points  in the redshift range 
$0.106 \lesssim z \lesssim 2.36$. The relevant $\chi^2$ function   reads as
\begin{equation}
    \label{chisq_BAOS}
    \chi^{2}_{BAO}\left(\phi^{\nu +1} \right) = 
\mathbf{s}C^{-1}_{cov}\mathbf{s}^{T} + 
\sum_{i=8}^{N}\frac{\left[T(z_i;\phi^{\nu}) - T^{obs}_{i}\right]^2
    }{\sigma_{i}^2},
\end{equation}
with $C^{-1}_{cov}$   the inverse of the covariance matrix  of the first 6 
measurements  available at \cite{Ryan:2019uor}. The vector $\mathbf{s}$ has as
elements the $s_{i}$, given as $s_{i} = d_{m} -d^{obs}_{i}r_d/r_{d,fid}$ for 
odd i and 
$s_{i} = H(z_{i};\phi^{\nu}) - H^{obs}_{i}r_{d,fid}/r_d$ for even i. In all 
cases, $r_{d,fid} = 147.78$. Furthermore, for $i \in \{8,9\}$,  
$T(z_i;\phi^{\nu}) 
= D_v(z_{i};\phi^{\nu})$, $ T^{obs}_{i} = D_{v,i}^{obs} r_d/r_{d,fid}$, with 
$r_{fid,8} = 148.69 \ Mpc$ and $r_{fid,9} = 147.66 \ Mpc$ respectively. For $i 
= 10$, $T(z_i;\phi^{\nu}) = cH(z_i;\phi^{\nu})^{-0.7} 
D_m(z_i;\phi^{\nu})^{0.3}/r_d$ and for 
$i=11$, $T(z_i;\phi^{\nu}) = cH(z_i;\phi^{\nu})^{-1}r_d$. 
Finally, in the expressions above the following quantities have been used
\begin{subequations}
    \begin{equation}
        D_M(z_i;\phi^{\nu}) = \frac{D_{L}(z_i;\phi^{\nu})}{1+z},
    \end{equation}
    \begin{equation}
        D_A(z_i;\phi^{\nu}) = \frac{D_{L}(z_i;\phi^{\nu})}{(1+z)^2},
    \end{equation}
    \begin{equation}
        D_V(z;\phi^\nu) = \left[\frac{cD_A(z;\phi^\nu)^2 \,z 
(1+z)^2}{H(z;\phi^\nu)}\right]^{1/3}\,.
    \end{equation}
    \end{subequations}

\subsubsection{Big Bang Nucleosynthesis}

Any cosmological scenario arising from modified gravity should preserve the 
standard thermal history of the universe. Hence, a basic and rough condition is 
applicable  in the form of an extra prior. Specifically, we 
require that the following inequality holds 
\cite{Torres:1997sn,Lambiase:2005kb,Barrow:2020kug}
\begin{equation}
    \frac{\left(H_{i}(z_{BBN};\phi^{\nu}) - H_{\Lambda }(z_{BBN};\Omega_{m0}) 
\right)^2}{H_{\Lambda}(z_{BBN};\Omega_{m0})^2} < 0.1,
\end{equation}
where $z_{BBN} \sim 10^9$. For the fiducial $\Lambda$CDM cosmology, namely 
$H_{\Lambda }$, we employ the parameter 
values from Planck  \cite{Aghanim:2018eyx}.

\subsubsection{Joint likelihood analysis}

      In order to  obtain the joint constraints on the cosmological parameters 
from  the aforementioned cosmological probes,
        we introduce the total likelihood function as  
        \begin{equation}
        \mathcal{L}_{\text{tot}}(\phi^{k}) =  \mathcal{L}_{SNIa}\times 
\mathcal{L}_{H}\times \mathcal{L}_{BAO}.
        \end{equation}
        It is easy to deduce that relevant  $\chi^2_{\text{tot}}$   is 
given as
        \begin{equation}
        \chi_{\text{tot}}^2(\phi^{k}) = \chi^{2}_{SNIa}+ \chi^{2}_{H}  + 
\chi^{2}_{BAO}. 
        \end{equation}
        The involved statistical vector has $k$ components, i.e. the $\nu$
        parameters of the scenario at hand plus   $\nu_{\text{hyp}}$
hyper-parameters from the imposed datasets, namely $k = \nu + 
\nu_{\text{hyp}}$. Hence, the vector containing the free parameters of the 
scenaria at hand    is 
$\phi^{k} = \{\Omega_{m0},h,\lambda,\mathcal{M},r_{d} \}$. Note however
that from a statistical 
point of view     there is no  distinction between the intrinsic 
hyper-parameters 
of a given dataset and the free parameters of a cosmological scenario.

Finally, for  the likelihood maximization  we use an 
affine-invariant 
Markov Chain Monte Carlo sampler \cite{AffineInvMCMC}, obtained in 
 the  Python package emcee \cite{emcee}. We use 1000 
chains (walkers) and 3500 
steps (states). As a prior  we employ firstly the conditions $0.0 <  
\Omega_{m0}  < 1$, $0.60 < h < 0.90$, $ -19.9 < M < -18.0$, $-0.9 < \lambda < 
2.8$, $135 < r_d < 160$, and secondly the BBN constraint described above.
Lastly, the convergence of the MCMC algorithm is verified with 
auto-correlation 
time implementation, and moreover for completeness the Gelman-Rubin criterion 
is calculated.

\subsection{Information Criteria and Model Selection}
\label{AICBIC}

As a last step we present the standard ways in order to   compare a set of 
cosmological scenarios, namely we apply  the Akaike Information Criterion (AIC)
\cite{Akaike1974}, the Bayesian Information Criterion (BIC) \cite{Schwarz1978}, 
and the Deviance 
Information Criterion \cite{Spiegelhalter2002}.  Moreover, we
present the standard $\chi^2_{\textrm{min}}/\textrm{dof}$, where ``dof'' stands 
for degrees of freedom,  usually defined as the number of the used data 
points minus the number of fitted parameters. In our case this gives dof=77.
Nevertheless, $\chi^2_{\textrm{min}}/\textrm{dof}$ should be used for 
illustrative purposes, as the degrees of freedom might be ambiguous for 
non-linear (in terms of the free parameters) models. \cite{Andrae:2010gh}. 

The AIC criterion is based on information theory  and it is an asymptotically 
unbiased estimator of the 
Kullback-Leibler information.
Under the standard assumption of 
Gaussian 
errors, the corresponding estimator for the AIC  criterion reads
\cite{Ann2002,Ann2002b}
\begin{equation}
 \text{AIC}=-2\ln(\mathcal{L}_{\text{max}})+2k+
 \frac{2k(k+1)}{N_{\text tot}-k-1}\,,
 \end{equation}
with $\mathcal{L}_{\text{max}}$   the maximum likelihood of the dataset(s) 
under consideration and $N_{\text tot}$   the total  data points  number.
It is apparent that for $N_{\text tot} >> 1$ this expression
 gives the original AIC version, namely $\text{AIC}\simeq 
-2\ln(\mathcal{L}_{\text{max}})+2k$. As it is discussed in
\cite{Liddle:2007fy}, it is considered as best practise to use the 
modified AIC criterion.
        
The BIC criterion is a Bayesian evidence   estimator, and it is written as
\cite{Ann2002,Ann2002b,Liddle:2007fy} 
\begin{equation}
\text{BIC} = -2\ln(\mathcal{L}_{\text{max}})+k \,{\text log}(N_{\text{tot}})\,.
\end{equation}
Finally, the DIC criterion employs both Bayesian statistics 
and 
information theory concepts \cite{Spiegelhalter2002}, and it is expressed as 
\cite{Liddle:2007fy}
\begin{equation}
{\text {DIC}} = D(\overline{\phi^k}) + 2C_{B}.
\end{equation}
The quantity   $C_{B}$ is the  Bayesian complexity 
$C_{B} = \overline{D(\phi^k)} - D(\overline{\phi^k})$,
where   overlines imply the standard mean value.  
Moreover, $D(\phi^k)$ is the Bayesian Deviation, which can be expressed as 
$D(\phi^k) = 
-2\ln[\mathcal{L}(\phi^k)]$ in the case of 
exponential
class 
of 
distributions. It is closely related to the number of effective 
degrees of freedom  \cite{Spiegelhalter2002}, which  is actually  the 
number of parameters that   affect  the fitting. In a less strict 
manner, it could be considered as a measure of the ``spread" of the likelihood. 

In contrast with AIC and BIC criteria, instead of using just the best fit 
likelihood, DIC uses the whole sample. Furthermore, AIC and BIC count and 
penalize all the involved parameters, while DIC penalizes only the number of 
parameters that  contribute to the fit in an actual way. Finally, an 
additional  appealing feature of DIC criterion is that its calculation is 
computationally light under the    MCMC samples.

Given a set of scenarios that describe the same class of phenomena, our problem 
is to sort the models according to their fitting efficiency in the context of 
the available data. We employ the aforementioned  three information 
criteria (IC)  and 
we calculate the relative difference of the IC value for the given set
of models, $\Delta 
\text{IC}_{\text{model}}=\text{IC}_{\text{model}}-\text{IC}_{\text{min}}$,
where the $\text{IC}_{\text{min}}$ is the minimum $\text{IC}$ value inside the 
 competing models set.
In order to qualify each model  in terms of its relevant adequacy, we apply 
the 
Jeffreys scale 
\cite{Kass:1995loi}. Specifically, 
the condition $\Delta\text{IC}\leq 2$  implies  statistical compatibility 
with the most favoured model by the data, while the condition 
$2<\Delta\text{IC}<6$ corresponds to middle tension between the two models, and 
lastly the condition $\Delta\text{IC}\geq 10$ implies strong tension.

\section{Results}
\label{results}

\begin{figure}[ht]
\includegraphics[scale=0.53]{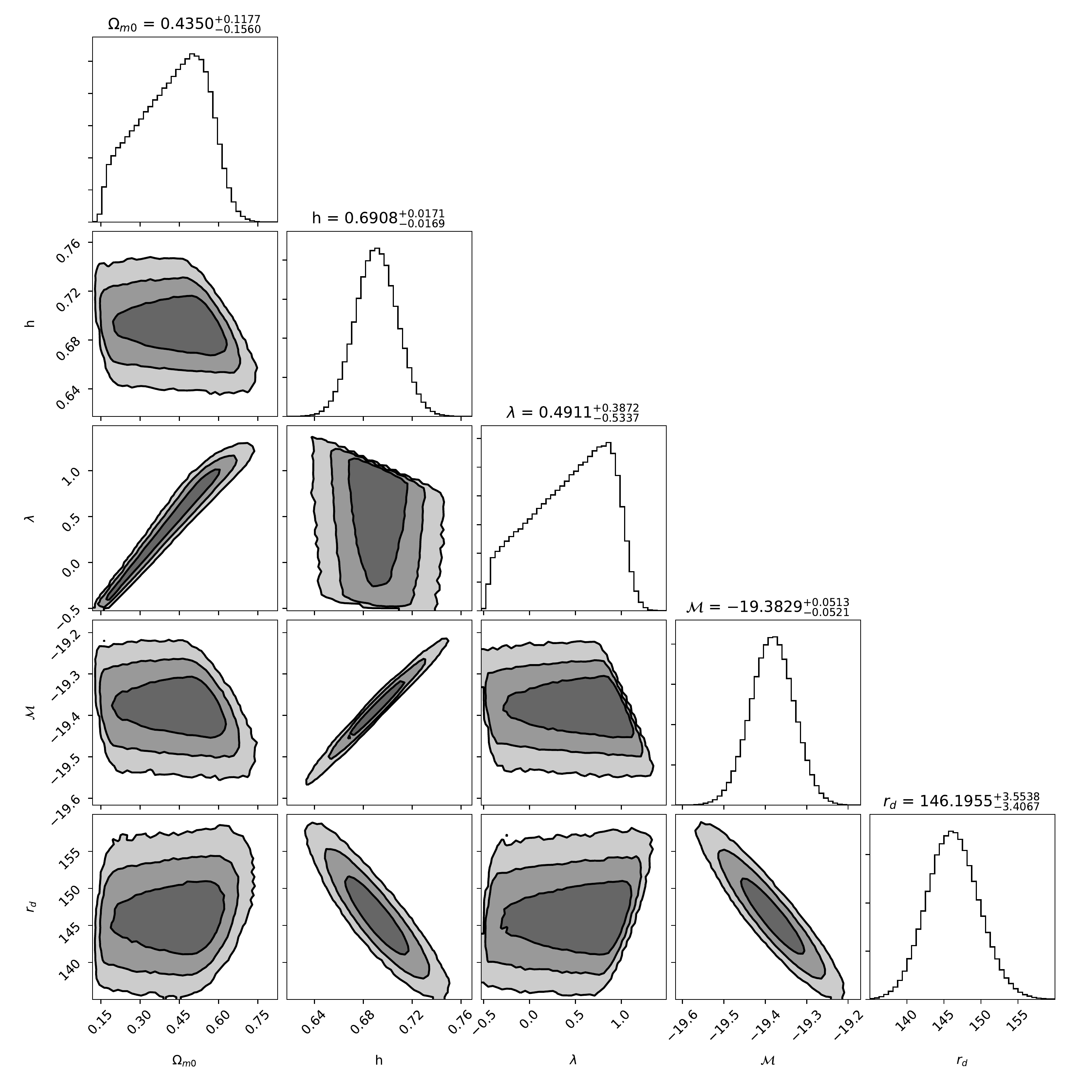}
\begin{center}
\caption{{\it{The $1\sigma$, $2\sigma$ and $3\sigma$ likelihood contours for
Model 1 of (\ref{rhoDEa}),(\ref{pDEa}), for all possible 2D subsets of the 
parameter space
$(\Omega_{m0},h,\lambda,\mathcal{M},r_{d})$. Moreover, we present the mean 
parameter
values   within   the $1\sigma$ area of the MCMC chain. We have 
performed a joint 
analysis of CC+SNIa+BAO  data.}}}
\label{figmod1}
\end{center}
\end{figure}

\begin{figure} [ht]
\centering\includegraphics[scale=0.53]{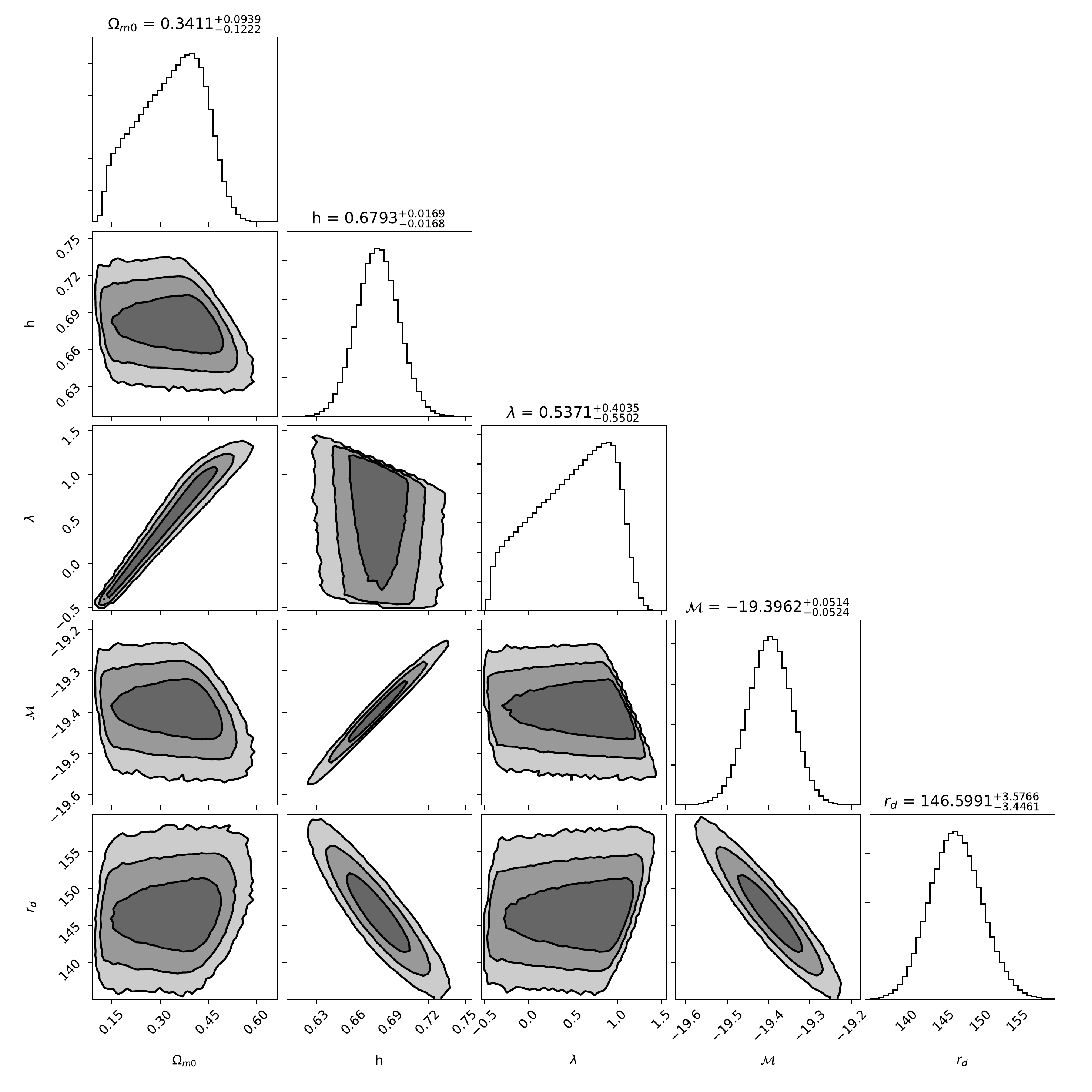}
\caption{{\it{The $1\sigma$, $2\sigma$ and $3\sigma$ likelihood contours for
Model 2 of (\ref{rhoDEb}),(\ref{pDEb}), for all possible 2D subsets of the 
parameter space
$(\Omega_{m0},h,\lambda,\mathcal{M},r_{d})$. Moreover, we present the mean 
parameter
values   within   the $1\sigma$ area of the MCMC chain. We have 
performed a joint 
analysis of CC+SNIa+BAO data.}}}
\label{figmod2}
\end{figure}

In this section we proceed to the observational analysis of   Myrzakulov 
gravity using the datasets and the methods described above.   Note that the 
free parameters of the aforementioned models are 
$\Omega_{m0},\ h$ and $\lambda$, while for the case of the concordance cosmology 
they are $\Omega_{m0},\ h$. 
For convenience we summarize the obtained results   in Table 
\ref{tab:Results1}. 
Additionally, in Figs. \ref{figmod1} and  \ref{figmod2} we present the 
corresponding  contour plots for Model 1 and Model 2 respectively. For 
comparison and benchmark we also analyzed the concordance model, namely 
$\Lambda$CDM. 
 
\begin{table}[!]
\tabcolsep 4.pt
\vspace{1mm}
\begin{tabular}{cccccccc} \hline \hline
Model & $\Omega_{m0}$ & $h$ & $\lambda $  & $r_{d}$ & $\mathcal{M}$  & 
$\chi_{\text{min}}^{2} $&  $\chi^2_{\text{min}}/dof$  \vspace{0.05cm}\\ 
\hline
\hline
 Mod. 1 & $0.425_{-0.146}^{+0.107}   $ & $0.691_{-0.017}^{+0.016}  $ & 
$0.491_{-0.533}^{+0.387}$ & $ 146.20_{-3.41}^{+2.55} $ &$ 
-19.382_{-0.052}^{+0.051}$
& 61.93& 0.8043 \\ 
 
 Mod. 2 & $ 0.339_{-0.122}^{+0.093}  $ & $0.679_{-0.016}^{+0.016}  $ & $  
0.537_{-0.550}^{+0.403} $& $146.60 _{-3.44}^{+3.57}$ &$ 
-19.396_{-0.052}^{+0.051} $ & 63.53 & 0.8251 
\\

$\Lambda$CDM & $0.292_{-0.014}^{+0.015}$ & $0.692 _{-0.017}^{+0.017} $& - & 
$145.87_{-3.38}^{+3.53}$ & $-19.377 
_{-0.052}^{+0.051}$ & 61.73 &  0.7914 \\  
 
\hline\hline
\end{tabular}
\caption[]{Observational constraints and the
corresponding $\chi^{2}_{\text {min}}$, as well as $\chi_{min}^2/dof$ 
(where 
``dof'' stands for degrees of freedom,  in our case dof = 77),
for the two Myrzakulov gravity models, 
presented previously, using CC/Pantheon/BAO data-sets. In order to allow  
direct comparison, the concordance flat $\Lambda$CDM model is also analyzed, 
 giving results very similar with the corresponding ones of 
\cite{Cao:2021ldv}.}
\label{tab:Results1}
\end{table}

 
As we can see, according to the combined analysis of CC+SNIa+BAO data  
 we acquire  $\sim 1 \sigma$ compatibility in all cases. The 
dimensionless parameter $\lambda$ is constrained to an interval around 0, that 
includes $\Lambda$CDM paradigm, which was expected since as we discussed above 
a realistic modified gravity should be a small deviation from general 
relativity. Nevertheless, note that in both Model 1 and Model 2, the 
$\lambda$-contours are slightly shifted towards positive values.
We mention that having the likelihood contours for the parameter 
$\lambda$ allows us to extract the constraints on the parameter $c$ 
through  expression (\ref{FR2a1b}) for Model 1 and on the   parameter $c_1$ 
through (\ref{FR2a1b2}) for 
Model 2. In particular, for 1$\sigma$ region for Model 1 we obtain 
$c=1.550^{+0.828}_{-0.876}$, while for Model 2 we find 
$c_1=4.94^{+2.28}_{-2.75}$.

Concerning the values of $\Omega_{m0}$ we observe that Model 1 gives a rather 
large value, due to the degeneracy with $\lambda$, while for Model 2 this is 
not the case. Concerning the Hubble constant $h$, for Model 1 we find that   
$0.690_{-0.017}^{+0.016} 
 $, while for Model 2 we obtain  $0.679_{-0.016}^{+0.016}  $. This implies that 
the 
obtained values for the 
present Hubble parameter $H_0$ are in between the  Planck estimation $H_0=67.36 
\pm 0.54$ km/s/Mpc \cite{Aghanim:2018eyx}  and the local estimation    
$H_0=73.24 \pm 1.74$ km/s/Mpc \cite{Riess:2016jrr}, although  closer to 
the former.  In addition, the extracted $H_0$ value for both models is 
consistent with other astrophysical inferences of Hubble constant, i.e $H_0 = 
67.4_{-3.2}^{+4.1} \ km s^{-1}Mpc^{-1}$ \cite{Birrer:2020tax},
and  $H_0 = 69.6 \pm 2.5 \ km s^{-1}Mpc^{-1}$  \cite{Freedman:2020dne}.   
It is 
interesting to note that     results in this range have been supported for 
about a decade now, amongst others in 
\cite{Chen:2011ab,Rigault:2014kaa,Chen:2016uno,Zhang:2017aqn,
Dhawan:2017ywl,Fernandez-Arenas:2017isq,Blum:2020mgu}.

\begin{figure}[!]
\begin{center}
\includegraphics[scale=0.4]{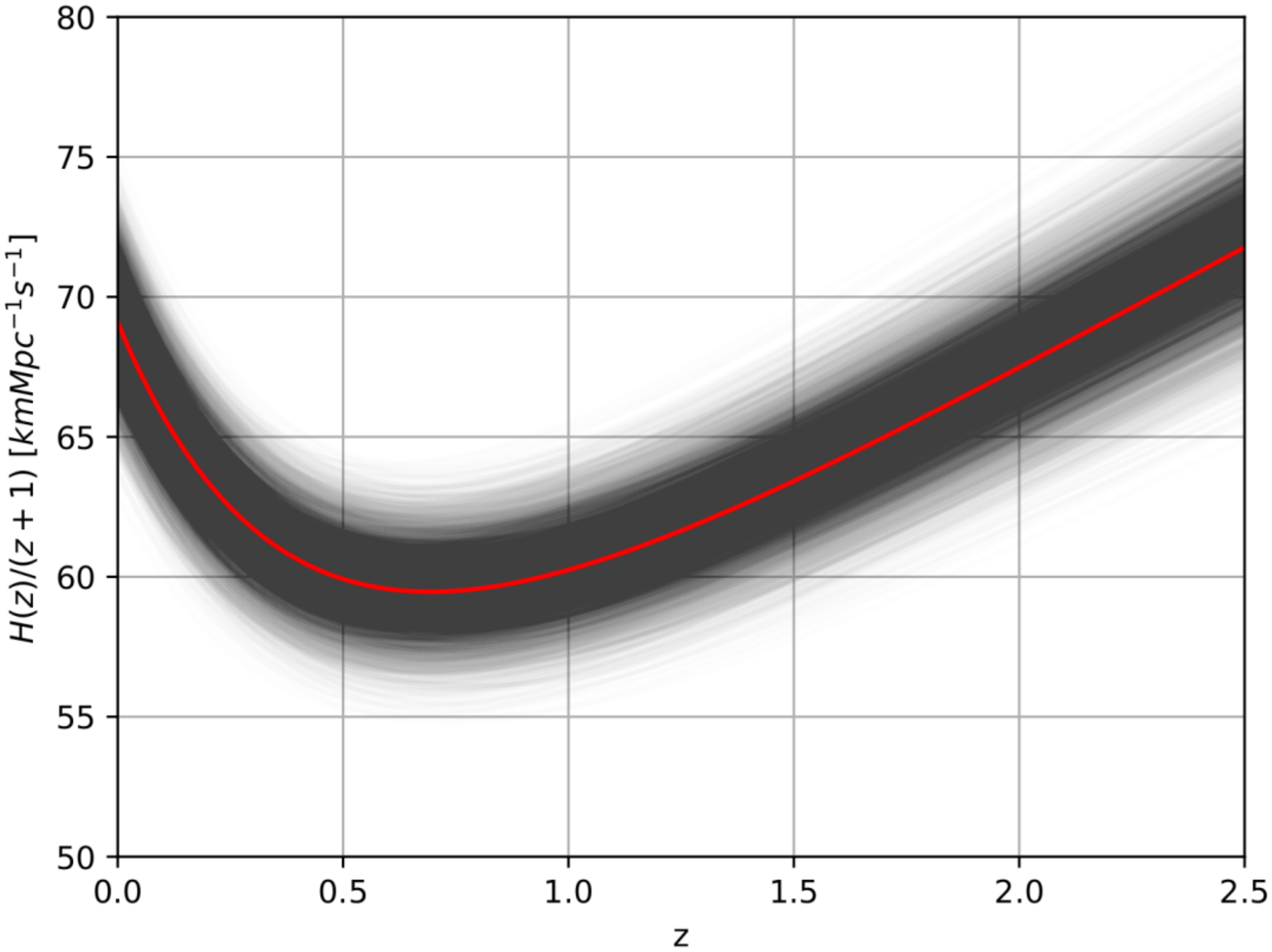}
\caption{{\it{
The reconstruction of the  $H(z)/(z+1)$ as a function of the redshift for Model 
1, arisen 
from (\ref{FR1a1}),(\ref{rhoDEa}). We re-sampled the chains produced by emcee  
taking 6000 samples, 
and we plot all the obtained   curves, alongside   the curve 
corresponding to the best fit of the parameters (red curve). 
}}}
\label{Hzmod1}
\end{center}
\end{figure}

\begin{figure}[!]
\begin{center}
\includegraphics[scale=0.4]{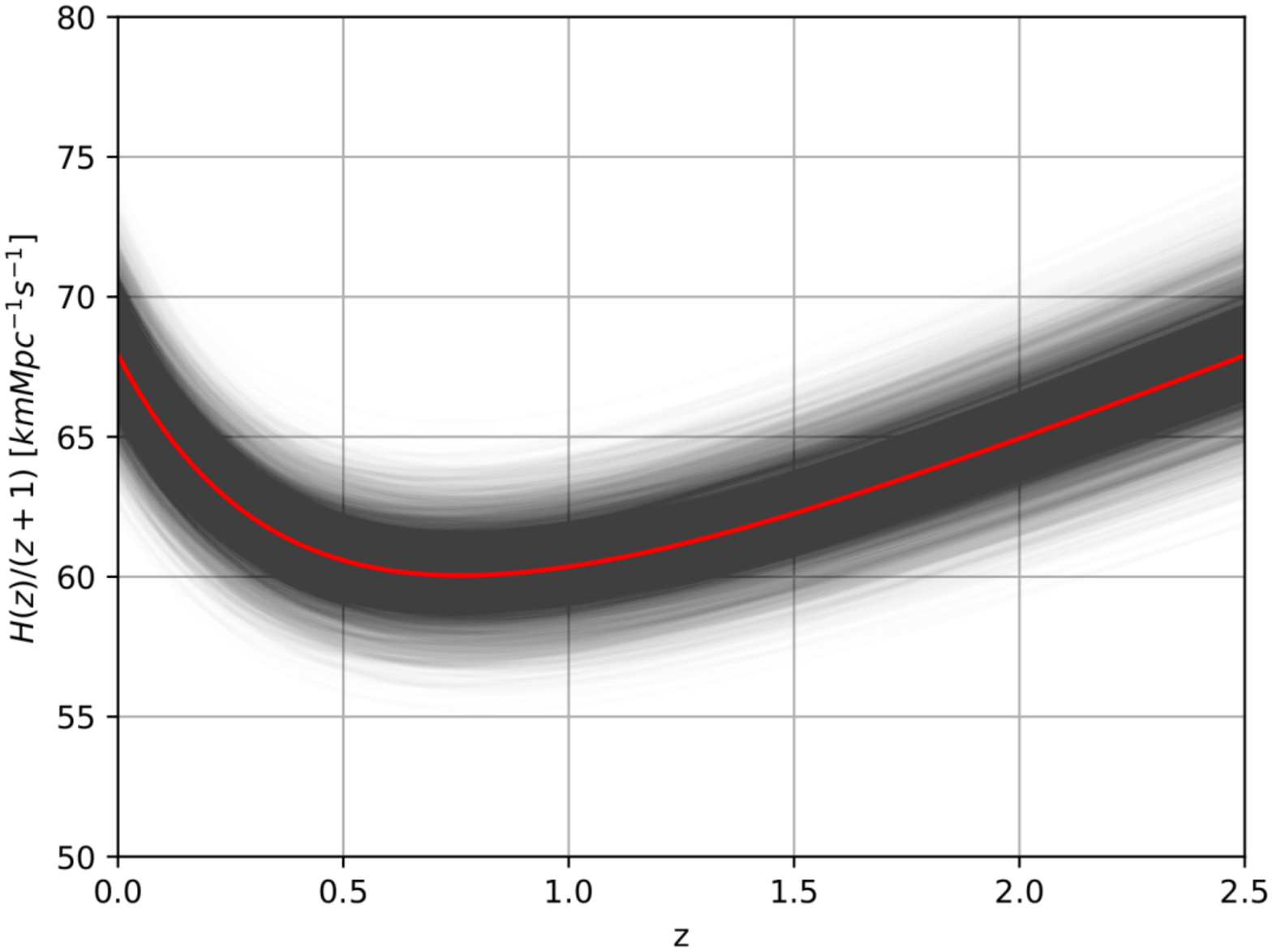}
\caption{{\it{
The reconstruction of the  $H(z)/(z+1)$ as a function of the redshift for Model 
2, arisen 
from (\ref{FR1a2}),(\ref{rhoDEb}). We re-sampled the chains produced by emcee  
taking 6000 samples, 
and we plot all the obtained   curves, alongside   the curve 
corresponding to the best fit of the parameters (red curve). 
}}}
\label{Hzmod2}
\end{center}
\end{figure}

In order to provide a more complete and transparent picture, 
we   use the obtained   allowed 
parameter values  in order to extract the resulting
$H(z)$. In Figs. \ref{Hzmod1} and \ref{Hzmod2} we present   the reconstructed 
mean  
$H(z)/(z+1)$ as a function of the redshift,  alongside the allowed curves for 
the 
1$\sigma$ allowed model parameters presented above, 
for Model 1 and Model 2  respectively. 
These graphs are quite similar with the corresponding ones for quintessence 
models ($\phi CDM$) of \cite{Farooq:2016zwm}.

 \begin{figure}[ht]
\begin{center}
\includegraphics[scale=0.44]{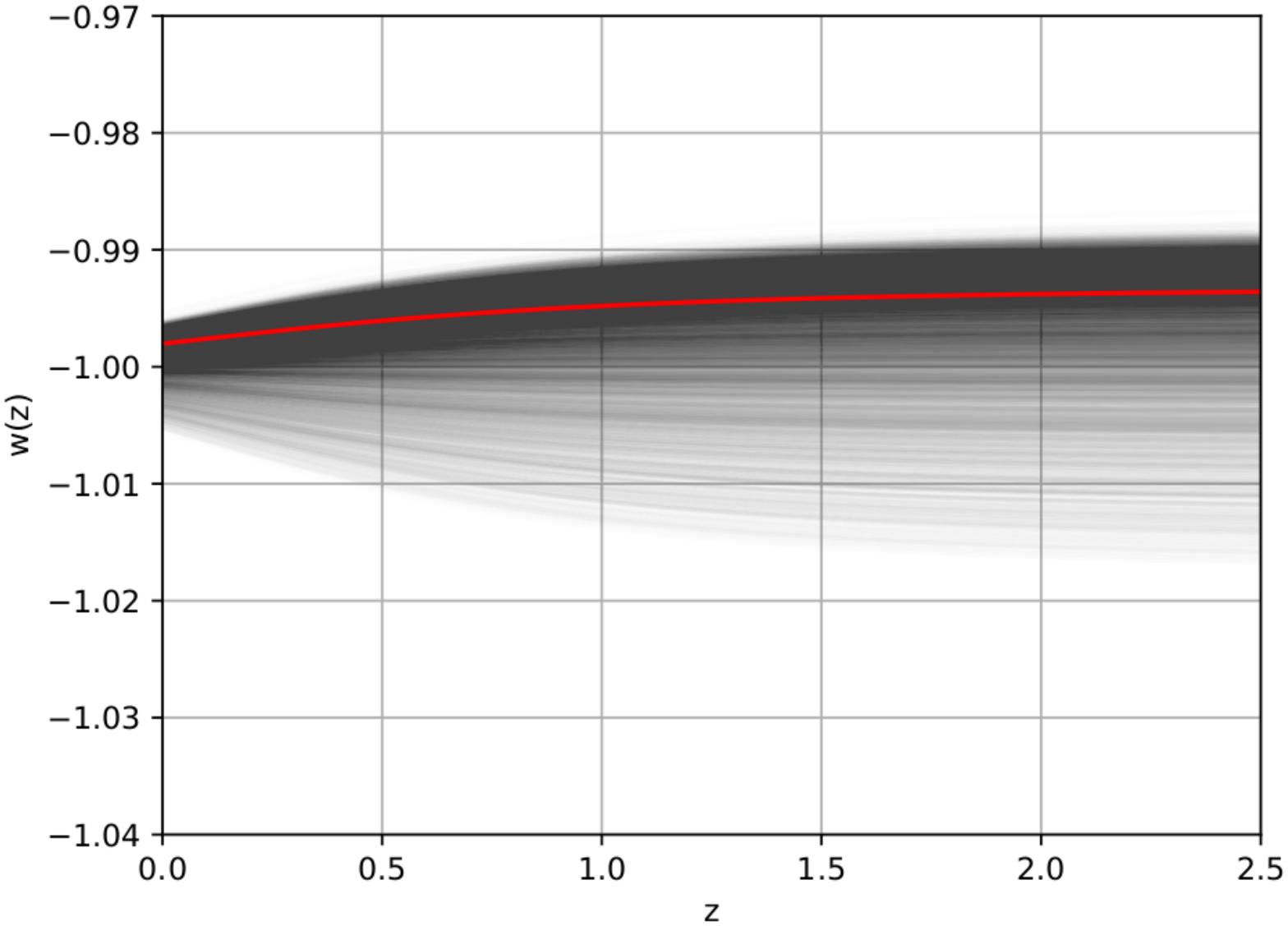}
\caption{{\it{The reconstruction of the effective dark-energy equation-of-state 
parameter $w_{de}(z)$ as a function of the redshift for Model 1 given by 
\eqref{wDEa}. We re-sampled the chains produced by emcee  taking 6000 samples, 
and we plot all the obtained $w_{de}(z)$ curves, alongside   the curve 
corresponding to the best fit of the parameters (red curve).}}}
\label{w_z_plotFor_Ml1}
\end{center}
\end{figure}

As a next step  we  investigate of the evolution of the dark energy 
equation-of-state parameter. In particular, having obtained the allowed 
parameter values at 1$\sigma$ confidence level, we can use them in order to 
extract the resulting $w_{de}(z)$ behavior given by (\ref{wde11}), with the 
deceleration parameter given by (\ref{qz_myrz1}) for Model 1 and by
(\ref{qz_myrz2}) for Model 2. 

\begin{figure}[ht]
\begin{center}
\includegraphics[scale=0.44]{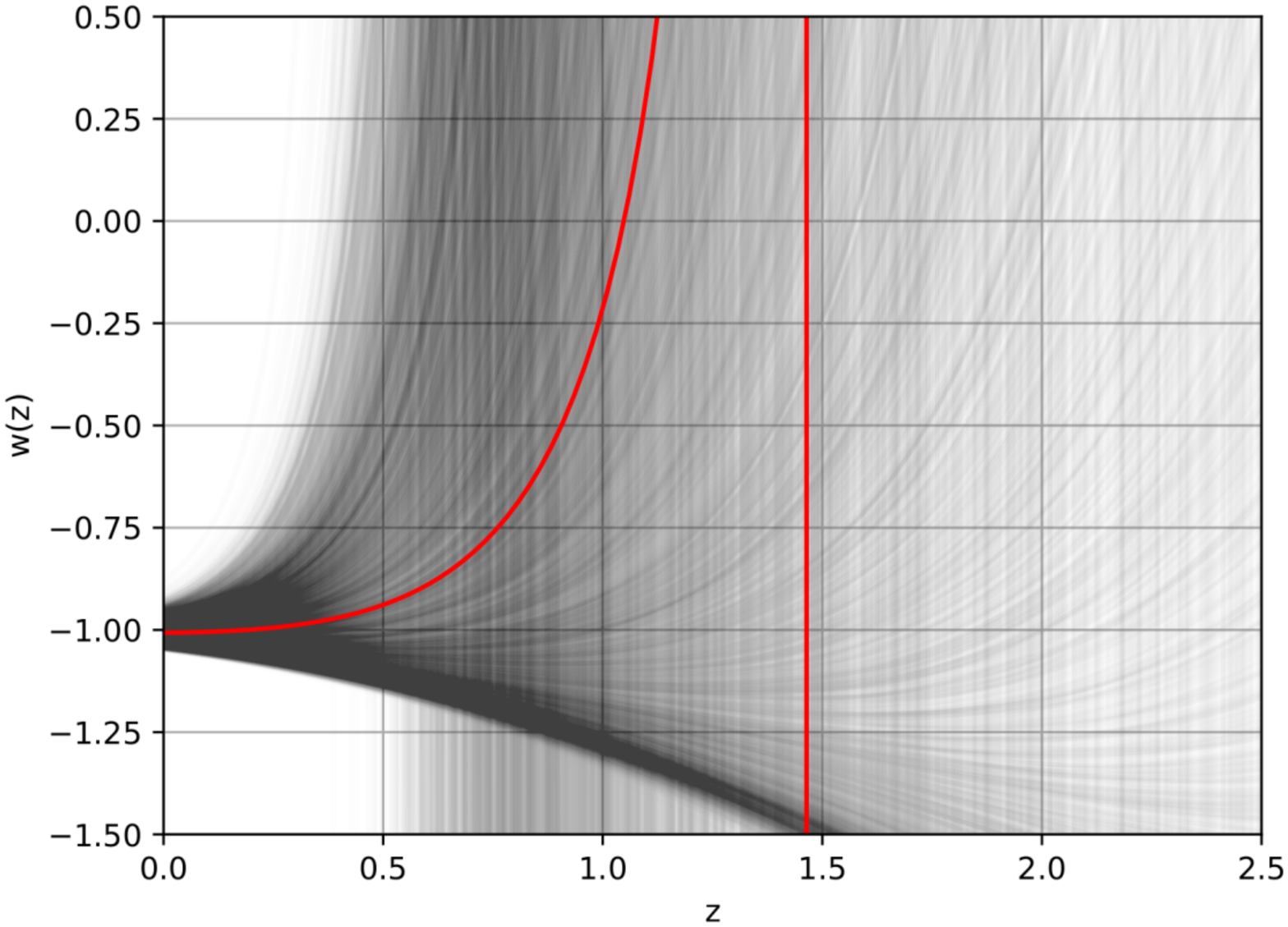}
\caption{{\it{
The reconstruction of the effective dark-energy equation-of-state 
parameter $w_{de}(z)$ as a function of the redshift for Model 2 given by 
\eqref{wDEb}. We re-sampled the chains produced by emcee  taking 6000 samples, 
and we plot all the obtained $w_{de}(z)$ curves, alongside   the curve 
corresponding to the best fit of the parameters (red curve).  The 
over-populated 
area at the bottom corresponds to a peak within 1$\sigma$ area, nevertheless 
since we extract the median value of each parameter within 1$\sigma$   as 
the best fit, the  ``best''  $w_{de}(z)$ curve differs.}}}
\label{w_z_plotFor_Ml2}
\end{center}
\end{figure}

In Fig.  \ref{w_z_plotFor_Ml1}  we depict the reconstructed mean   
$w_{de}(z)$ (red curve) for Model 1, alongside the allowed curves for the 
1$\sigma$ allowed model parameters presented above. As we observe, the 
corresponding behavior is very close to $\Lambda$CDM scenario for every 
parameter values. Similarly, in Fig.  \ref{w_z_plotFor_Ml2}  we present the 
corresponding graph for Model 2. In this case the scenario resembles  
$\Lambda$CDM at low redshifts, however at earlier times the mean 
curve, as well as  many of the ``individual'' curves, present a deviation, since 
this is allowed by the used datasets.
In  particular, for some parameter choices the dark-energy pressure at a 
particular redshift diverges and changes 
sign, and thus the $w_{de}(z)$ transits on the other side of the phantom-divide. 
Such   
 energy conditions violations are common in modified gravity theories, and 
actually
they can lead to interesting cosmological phenomenology.
Note   that the observable quantities (the Hubble function and its derivatives, 
the density parameters etc) remain 
finite. However, we mention that a significant sub-set of the curves, i.e. 
  a large region of the parameter space of the model, does not exhibit such a 
behavior,
and the individual obtained 
curves resemble the  $\Lambda$CDM evolution.
 
Furthermore,  we proceed to the reconstruction of the decceleration parameter 
using random sampling of 
the obtained chains. Concerning the current value $q_0$,
for Model 1 using 
\eqref{q_0_myrz1}  we obtain $q_0 = 
-0.561_{-0.021}^{+0.022} $, while for Model 2 using \eqref{q_0_myrz2} 
we acquire $q_0 = -0.880_{-0.009}^{+0.010}$. These are in   agreement 
with the   values obtained using other datasets, such as supernovae, quasars and
gamma-ray bursts   by means of model-independent techniques 
\cite{Rezaei:2020lfy}.

 Additionally, we calculate the transition redshift,  for the two 
models, using relations
  (\ref{ztrmod1}) and  (\ref{ztrmod2}) respectively.
 For Model 1 we find $z_{\textrm{tr},1}=0.36_{-0.18}^{+0.10}$, while
 for Model 2 we acquire $z_{\textrm{tr},2}=0.74_{-0.14}^{+0.07}$. 
 It is of interest to compare the aforementioned values with $z_{\textrm{tr},A} 
= 0.72 \pm 0.05$, 
 \cite{Farooq:2016zwm} and $z_{\textrm{tr},B} =
0.64^{+0.12}_{-0.09}$, \cite{Haridasu:2018gqm}.
For Model 1   we observe mild compatibility within $\sim 3.5\sigma$ and within 
$\sim 3\sigma$ with ``A'' and ``B'' results, respectively. On the other hand, 
for the case of Model 2  we report 1 $\sigma$ compatibility  with both results. 
These results act as an additional verification check of the examined models.

We close this analysis with the examination of
 the statistical significance of our fitting
results, applying the AIC, BIC and DIC information criteria described in 
subsection \ref{AICBIC}. We summarize our 
 results in Table \ref{tab:Results2}. As we observe,   Model 1 is statistically 
equivalent with $\Lambda$CDM paradigm, and especially the combined and more 
complete  DIC criterion gives an almost equal value. Additionally, Model 2 also
presents  a very good fitting behavior, and according to DIC it is also 
statistically equivalent with $\Lambda$CDM paradigm, which is an interesting 
result since Model 2 does not contain $\Lambda$CDM scenario as a limit for any 
parameter value.
\begin{table}[ht]
\begin{center}
\tabcolsep 4.0pt
\vspace{1mm}
\begin{tabular}{ccccccc} \hline \hline
Model & AIC & $\Delta$AIC & BIC &$\Delta$BIC & DIC & $\Delta$DIC
 \vspace{0.05cm}\\ \hline
\hline
 Mod. 1 & 72.7234 & 2.4757 & 83.9675  & 4.6124 & 69.6728 & 0.0007 \\  
Mod. 2 & 74.3204  & 4.0727 &85.5645 & 6.2094 & 71.3725  & 1.7004 \\ 
$\Lambda$CDM & 70.2477  & 0 & 79.3551 & 0 & 69.6721  & 0  \\ 
\hline\hline
\end{tabular}
\caption{The information criteria 
AIC, BIC and DIC for the examined cosmological models, alongside
the relative difference from the best-fitted model $\Delta\text{IC} 
\equiv \text{IC} - \text{IC}_{\text{min}}$.
\label{tab:Results2}}
\end{center}
\end{table}

\section{Conclusions}
\label{Conclusions}

In this work we have used observational data from Supernovae (SNIa) Pantheon 
sample, from Baryonic Acoustic Oscillations (BAO), and from cosmic chronometers
  measurements of the Hubble parameter   (CC), alongside arguments from Big 
Bang Nucleosynthesis (BBN), in order to extract 
constraints on  Myrzakulov $F(R,T)$ gravity. This is a connection-based theory 
belonging to the  Riemann-Cartan subclass, 
that uses a specific but non-special 
connection, which then leads to  extra degrees 
of freedom. One introduces a parametrization 
  that quantifies the deviation of torsion and curvature    scalars 
form their   values corresponding to the special Levi-Civita 
and Weitzenb{\"{o}}ck  connections, and then constructs various models by 
assuming specific forms for the involved functions. In all models, one obtains 
an effective dark-energy sector of geometrical 
origin.

We considered two specific models, which are known to lead to interesting 
phenomenology. Our analysis shows that both models are capable of describing 
 adequately the imposed datasets, namely  CC+SNIa+BAO ones, obtaining $\sim 1 
\sigma$ compatibility in all cases. Concerning Model 1, which includes   
$\Lambda$CDM paradigm as a particular limit, we found a relatively large value 
for $\Omega_{m0}$ and a value for $h$ in between the Planck and local 
estimation,  although  closer to 
the former. For the  
dimensionless parameter $\lambda$ we found 
that it is constrained to an interval around 
0, which 
corresponds to $\Lambda$CDM scenario, however the 
corresponding contours are slightly shifted towards positive values.
In the case of Model 2, we found   smaller $\Omega_{m0}$ and   $h$, while 
$\lambda$ is again constrained around 0 with favoured positive values.

 Additionally, we used the obtained posterior distribution of the parameters 
at 
  1$\sigma$ confidence level, and we reconstructed the  Hubble function as a 
function of the redshift.
  As we showed, the obtained graphs for  $H(z)/(z+1)$ are in very good agreement 
with observations.
 Furthermore, we  reconstructed the induced dark-energy 
equation-of-state parameter as a function of the redshift. As we saw, for Model 
1 $w_{de}(z)$ is very close to $\Lambda$CDM scenario, while  for  Model 2 it  
resembles  $\Lambda$CDM at low redshifts, however at earlier times
 deviations  are  allowed.
  
  Finally,   applying the AIC, BIC and the combined DIC criteria, we deduced 
that both Model 1 and Model 2 present   a very efficient fitting behavior, and 
are statistically equivalent with $\Lambda$CDM cosmology. This is an 
interesting result since Model 2 does not contain $\Lambda$CDM scenario as a 
limit for any parameter value.
 
 In summary,  Myrzakulov $F(R,T)$ gravity   
is  in agreement with
cosmological data, and   it could serve as a   candidate for the description 
of nature. Nevertheless, one should also investigate the theory at the 
perturbation level and confront it with perturbation-related data, i.e 
$f\sigma_8$. Such an analysis, although  both interesting 
and necessary, lies beyond the scope of the present work and it is left for a 
future project.


\providecommand{\href}[2]{#2}\begingroup\raggedright\endgroup
\end{document}